\documentclass[9pt,shortpaper,twoside,web]{ieeecolor2}
\usepackage{generic}
\usepackage{cite}
\usepackage{amsmath,amssymb,amsfonts}
\usepackage{algorithmic}
\usepackage{graphicx}
\usepackage{textcomp}
\usepackage{booktabs}
\def\BibTeX{{\rm B\kern-.05em{\sc i\kern-.025em b}\kern-.08em
    T\kern-.1667em\lower.7ex\hbox{E}\kern-.125emX}}
\begin{document}
\title{Body Habitus Dominates Solver Choice as a Source of Uncertainty in MRI Safety Assessment of Active Implantable Medical Devices}
\author{Safa Hameed, Fuchang Jiang, Bhumi Bhusal, Sana Ullah, Pia Sanpitak, Laleh Golestanirad, \IEEEmembership{Member, IEEE}
\thanks{This work was supported by the National Institutes of Health through grants R01EB030324, R01HL168859, R01EB034377 and R01EB036272.}
\thanks{Safa Hameed is with the Department of Biomedical Engineering, Northwestern University, Evanston, IL 60208 USA, and also with the Department of Radiology, Feinberg School of Medicine, Northwestern University, Chicago, IL 60611 USA. (e-mail: safa.hameed@northwestern.edu). }
\thanks{Fuchang Jiang is with the Department of Biomedical Engineering, Northwestern University, Evanston, IL 60208 USA, and also with the Department of Radiology, Feinberg School of Medicine, Northwestern University, Chicago, IL 60611 USA. (e-mail: fuchang.jiang@northwestern.edu). }
\thanks{Bhumi Bhusal is with the Department of Radiology, Feinberg School of Medicine, Northwestern University, Chicago, IL 60611 USA. (e-mail: bhumi.bhusal@northwestern.edu).}
\thanks{Sana Ullah is with the Department of Radiology, Feinberg School of Medicine, Northwestern University, Chicago, IL 60611 USA. (e-mail: sana.ullah@northwestern.edu). }
\thanks{Pia Sanpitak is with the Department of Biomedical Engineering, Northwestern University, Evanston, IL 60208 USA, and also with the Department of Radiology, Feinberg School of Medicine, Northwestern University, Chicago, IL 60611 USA. (e-mail: pia.sanpitak@northwestern.edu). 
}
\thanks{Corresponding Author: Laleh Golestanirad is with the Department of Biomedical Engineering, Northwestern University, Evanston, IL 60208 USA, and also with the Department of Radiology, Feinberg School of Medicine, Northwestern University, Chicago, IL 60611 USA. (e-mail: laleh.rad1@northwestern.edu).}}

\maketitle

\begin{abstract}
\textcolor{cyan}{\textit{Objective:}}
MRI is increasingly critical for patients with active implantable medical devices (AIMDs), yet access depends on safety labeling derived from computational heating predictions under ISO/TS 10974 Tier~3. Published assessments have relied predominantly on a single electromagnetic solver class and one or two standard-BMI reference anatomies, leaving the relative contributions of solver choice, tissue property uncertainty, and patient anatomy to predictive variability uncharacterized within a common workflow. \textcolor{cyan}{\textit{Methods:}} We performed a cross-platform evaluation of finite-difference time-domain (FDTD, Sim4Life) and finite element method (FEM, ANSYS HFSS) implementations of the full Tier~3 workflow for a deep brain stimulation system at 1.5 T, extending the analysis across more than 250 clinically realistic trajectories spanning standard male and female references (Duke, HBM, Ella), an elderly male (Glenn), and elevated-BMI models of both sexes (Fats, Ella BMI\,30). \textcolor{cyan}{\textit{Results:}} FDTD and FEM agreed closely in standard anatomies, with Maximum Allowable $B_1^+$ limits converging near 2.6-3.0\,$\mu$T. The elderly male model produced a comparable limit to Duke, indicating BMI rather than age drives heating variability. Elevated BMI reduced safe $B_1^+$ by 19-31\% in both sexes, while sex at matched BMI had no significant effect. Geometric morphing approximated the native obese limit, whereas dielectric property sweeps failed to reproduce elevated-BMI heating distributions. \textcolor{cyan}{\textit{Conclusion:}} Body habitus is the dominant source of predictive uncertainty in Tier~3 assessment, exceeding solver choice, dielectric assumptions, and sex. \textit{Significance:} Anatomical diversity, including elevated-BMI female phenotypes, should be treated as a primary variable. 
\end{abstract}

\begin{IEEEkeywords}
MRI safety, RF-induced heating, active implantable medical devices, deep brain stimulation, ISO/TS 10974, transfer function, anthropometric variability, geometric morphing, electromagnetic simulation.
\end{IEEEkeywords}

\section{Introduction}
\label{sec:introduction}
Magnetic resonance imaging (MRI) is a cornerstone modality across medicine, yet access to MRI remains constrained for many patients with active implantable medical devices (AIMDs) because MRI electromagnetic fields can interact with conductive implant components and create hazards, most notably RF-induced heating \cite{Nordbeck2008,Bhusal2018,Jiang2023EMBC,Brown2023,Kozlov2019,jiang2023age}. The clinical need is substantial: an estimated 75\% of patients with cardiac implantable electronic devices (CIEDs) will require MRI during their device lifetime \cite{Kalin2005,Taruya2015}. The safety risks and need for rigorous electromagnetic modeling persist even for patients with abandoned or retained cardiac leads, a population frequently denied MRI access due to unpredictable RF absorption profiles across different imaging landmarks \cite{Golestanirad2019a, Nguyen2022}. Similarly, 70\% of individuals with deep brain stimulation (DBS) systems require MRI within 10 years of implantation \cite{Falowski2016,Boutet2020}. When MRI access is denied, patients are often routed to ionizing-radiation-based alternatives; studies in pediatric implanted-device populations report up to a four-fold increase in cumulative effective radiation dose \cite{Aboyewa2024}.

The primary mechanism generating RF-induced heating in AIMDs is RF coupling: the incident RF electric field can couple to elongated conductive structures (e.g., leads and extensions), inducing currents that concentrate power deposition in surrounding tissue and produce localized temperature rise, often near electrode contacts or other discontinuities \cite{Nordbeck2008,Bhusal2018}. In DBS systems, this coupling mechanism is highly sensitive to the extracranial trajectory of the implanted leads. Recent clinical and phantom studies have demonstrated that strategically modifying the surgical routing of DBS leads (for example, adding concentric loops near the burr hole) can alter the orientation of the leads relative to the incident electric field, reducing RF-induced heating by up to three-fold at 1.5 T and 3 T \cite{Golestanirad2019b, Vu2024}. While this mechanism is well understood, expanding MRI access is often bottlenecked by the \emph{cost} and \emph{reproducibility} of generating safety evidence that supports scanner-operational exposure limits (e.g., expressed in terms of $B_{1}^{+}$ or related RF metrics). 

High-fidelity RF heating evaluations can require specialized commercial EM solvers, curated anatomical model libraries, and extensive measurement campaigns, which can be burdensome for small and mid-size manufacturers. While recent literature has explored the use of deep learning and machine learning algorithms to rapidly predict local specific absorption rates at lead tips and alleviate these computational bottlenecks \cite{Vu2021, Chen2023}, regulatory-facing safety assessments still rely predominantly on high-fidelity, full-wave numerical EM simulations. As a result, otherwise feasible MRI access expansions may be delayed or never pursued because results can be toolchain-dependent across organizations and because phenotype-diverse anatomical coverage is often prohibitively expensive.

A widely adopted international framework for AIMD MRI safety assessment is ISO/TS 10974 (``Assessment of the safety of magnetic resonance imaging for patients with an active implantable medical device'') \cite{ISO10974}. In particular, Clause~8 outlines a framework to predict RF-induced heating in realistic body configurations and is widely used in regulatory-facing MRI safety practice. The U.S. Food and Drug Administration lists ISO/TS 10974 among recognized consensus standards relevant to MRI safety and references it within MRI labeling/testing guidance for devices where MRI electromagnetic fields may affect safety or performance \cite{ISO10974,FDA_AccessData10974}. 

Within ISO/TS 10974, RF heating evaluation is organized into progressively more realistic assessment ``tiers.'' For elongated AIMDs, the highest-fidelity pathway in Clause~8, commonly referred to as Tier~3, combines computational EM field modeling in anatomically realistic scenarios with device-specific response characterization, most commonly implemented via the transfer function (TF) method, enabling prediction of RF-induced temperature rise across clinically relevant configurations without repeating full experimental heating tests for every scenario \cite{ISO10974,Park2007,Zheng2020,Feng2015,Mattei2021,Missoffe2018}. Recent advances that improve TF acquisition and estimation have further streamlined the device-characterization phase \cite{Long2024,Stijnman2022,Eijbersen2024}. Confidence in Tier~3 predictions depends on computational modeling choices (e.g., solver methodology and assumed tissue dielectric properties) and on the representativeness of anatomical models used to simulate \textit{in vivo} exposure \cite{Nguyen2020,Bhusal2020,BhusalPMC,Bottauscio2024,Carluccio2021}. Although these factors are recognized contributors to variability, their \emph{relative} contributions have not been systematically quantified in the context of the complete Tier~3 workflow, limiting guidance on where resources should be invested to obtain conservative, reproducible safety limits.
\begin{figure}[!t]
\centering
\includegraphics[width=\linewidth]{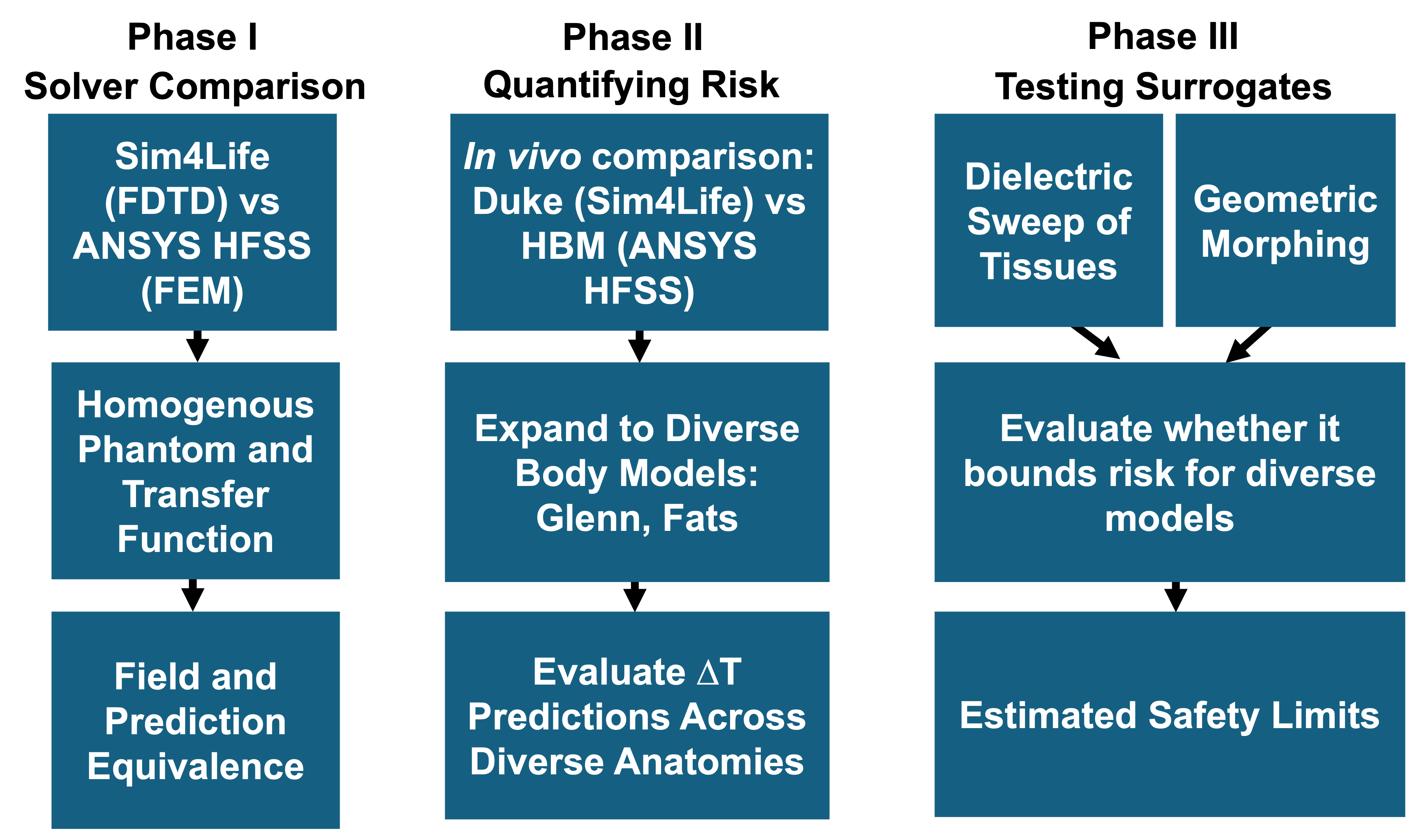}
\caption{Schematic overview of the study design and computational workflow. 
Phase~I (Solver Comparison) establishes cross-platform equivalence between  FDTD (Sim4Life) and FEM (ANSYS HFSS) solvers using a homogeneous phantom and transfer function (TF)-based predictions. 
Phase~II (Quantifying Risk) transitions to \textit{in vivo} assessments, 
first confirming equivalence in standard reference models (Duke, HBM), 
and evaluating the shift in heating distributions with 
diverse anatomical phenotypes (Glenn, Fats, Ella, Ella BMI~30).
Phase III (Testing Surrogates) evaluates two distinct strategies—dielectric tissue sweeps and geometric morphing—to determine which method better approximates safety risks and estimates $B_1^+$ limits for anatomically diverse populations.}
\label{fig:fig1}
\end{figure}

Two practical barriers limit the interpretability, portability, and broad adoption of Tier~3 (EM+TF) workflows, particularly for cost-effective MRI labeling evidence. First, solver dependence remains unresolved. Computational MRI safety studies have predominantly employed finite-difference time-domain (FDTD) solvers, in part due to their straightforward integration with widely used voxel-based anatomical libraries \cite{Kozlov2019,Xu2014}. Commercial platforms such as Sim4Life provide dedicated pipelines (e.g., IMAnalytics) that operationalize FDTD-based TF assessments \cite{sim4life_imanalytics}. In contrast, finite element method (FEM) solvers enable conformal meshing and local refinement, mitigating staircasing artifacts inherent to voxelized discretizations \cite{Kozlov2019,Makarov2017}. Despite widespread use of both paradigms, the literature lacks a systematic head-to-head cross-platform comparison of FDTD and FEM implementations for the TF Tier~3 workflow, from field calculations through \textit{in vivo} safety limits \cite{Kozlov2019}.

Second, anatomical diversity is expensive, yet it can dominate the uncertainty budget. Reliance on a single reference anatomical model (e.g., Duke) may underestimate risk for patients with elevated body mass index (BMI) \cite{Neufeld2011}. While high-BMI models exist in some libraries, differences in model representation (voxel vs.\ surface/mesh) can limit cross-platform applicability, and acquiring multiple phenotype-specific models increases cost and complexity \cite{Kozlov2019,Xu2014,Makarov2017,Neufeld2011}. A fundamental question is whether simple geometric transformations of a single model can approximate heating profiles observed in native phenotype models \cite{Makarov2017,Christ2010,Gosselin2014}. Geometric morphing has been used in other dosimetric domains (e.g., CT imaging) to generate high-BMI models \cite{Ding2012}, and scaling approaches have been explored in MRI safety research \cite{Jabbarigargari2025}. However, the validity of geometric morphing as a surrogate for native high-BMI models in AIMD safety assessment has not been established, nor is it clear whether more resource-intensive alternatives such as comprehensive dielectric property sweeps are required.

In this study, we address these sources of uncertainty through a systematic evaluation of the Tier~3 workflow, extending a preliminary conference abstract \cite{Hameed2025ISMRM}. Our objectives were to: (1) establish cross-platform equivalence between FDTD (Sim4Life) and FEM (ANSYS HFSS) solvers, from field-level validation through TF prediction to \textit{in vivo} safety limits; (2) quantify the impact of anatomical phenotype by comparing standard reference models against high-BMI anatomy across clinically realistic trajectories; and (3) evaluate whether practical surrogate strategies---dielectric property variation or geometric morphing---can approximate the elevated risk observed in high-BMI phenotypes. To align with best practices for method comparison in medical imaging, we treat cross-solver assessment as an equivalence problem rather than a null-hypothesis ``difference'' test, applying two one-sided tests (TOST) with pre-specified equivalence margins and Bland--Altman analysis to characterize solver-to-solver bias in TF-based predictions. To strengthen generalizability across patient populations, we include both male and female anatomical models in addition to phenotype variants such as high-BMI body habitus. Because experimental validation is central to interpreting equivalence, we further estimated the experimental uncertainty of the heating validation measurements (run-to-run repeatability and probe placement sensitivity) and used these results to contextualize agreement metrics and justify equivalence margins.

We also aim to make the workflow reproducible and low-cost for other groups. We therefore commit to sharing the complete analysis pipeline required to reproduce the reported results, including RF coil parameterization, TF processing and calibration code, post-processing routines for heating distributions and Maximum Allowable $B_{1}^{+}$ derivation, and statistical analysis code (TOST and bootstrap resampling). We will provide de-identified trajectory geometries and derived summary datasets sufficient to reproduce all figures and tables. Where redistribution of vendor-licensed assets is restricted, we will provide documented build instructions and solver-agnostic intermediate outputs (e.g., incident fields sampled on common grids and TF inputs) so that the full inference chain remains auditable and reproducible.
\begin{table}[t]
\centering
\caption{Computational anatomical models used in this study.}
\label{tab:anatomical_models}

\small
\resizebox{\columnwidth}{!}{%
\begin{tabular}{lcccccc}
\toprule
\textbf{Model} & \textbf{Sex} & \textbf{Age} & \textbf{BMI} & \textbf{Representation} & \textbf{Solver} & \textbf{Source / Library} \\
\midrule
Duke & Male & 34 & 22.4 & Voxel & Sim4Life & ViP 3.0 \\
Glenn & Male & 84 & 24.1 & Voxel & Sim4Life & ViP 3.0 \\
Fats & Male & 37 & 36.2 & Voxel & Sim4Life & ViP 3.0 \\
Ella & Female & 26 & 21.6 & Voxel & Sim4Life & ViP 3.0 \\
Ella BMI 30 & Female & 26 & 30.0 & Voxel & Sim4Life & ViP 3.0 \\
HBM & Male & -- & -- & Surface mesh & ANSYS HFSS & ANSYS HBM \\
Morphed HBM & Male & -- & -- & Surface mesh & ANSYS HFSS & Derived from HBM \\
\bottomrule
\end{tabular}}
\end{table}
 \section{Methods}
This study comprised three sequential phases designed to identify and quantify the dominant sources of variation in ISO/TS~10974 Clause~8 Tier~3 safety assessments (Fig.~\ref{fig:fig1}). Importantly, Sim4Life and ANSYS~HFSS represent the two dominant full-wave discretization paradigms for Maxwell’s equations: FDTD and FEM, respectively. Phase~I performed a controlled, like-for-like cross-solver comparison of incident-field calculations by evaluating $B_1^+$ and electric-field distributions generated by matched RF body-coil models loaded with an identical homogeneous phantom. Phase~II validated the complete Tier~3 transfer-function (TF) workflow by comparing TF-based temperature-rise predictions derived from each solver’s incident fields against controlled MRI heating measurements for two electromagnetically distinct DBS configurations (lead-only and full system). Phase~III quantified anatomy-driven effects by performing \textit{in vivo} simulations across standardized reference anatomical models and diverse phenotypes, including an elderly male model (``Glenn''), an obese male model (``Fats''), a standard adult female model (``Ella''), and an elevated-BMI female model (``Ella BMI~30''), and by evaluating practical surrogate strategies (dielectric property variation and geometric morphing) to approximate elevated-risk loading without requiring additional native phenotype models (Table~\ref{tab:anatomical_models}).

\subsubsection{Computational Resources}
\label{sec:comp_resources}
All electromagnetic simulations were performed in Sim4Life V8.0 (Zurich MedTech, Zurich, Switzerland) using an FDTD solver, and in ANSYS HFSS 2021 (ANSYS Inc., Canonsburg, PA) using an FEM solver. Solver hardware, runtimes, and peak memory are reported in Supplementary Methods.

\subsection{RF Coil Modeling}\label{sec:rf_coil_modeling}
A high-pass birdcage body coil representative of a Siemens AERA 1.5~T MRI system was modeled in both simulation platforms based on manufacturer specifications. The coil was tuned to 63.6~MHz and driven in quadrature to generate a circularly polarized $B_1^+$ field. In both solvers, the coil was loaded with an identical homogeneous cylindrical phantom positioned at isocenter (diameter 41~cm, height 42~cm), with dielectric properties $\sigma = 0.50$~S/m and $\varepsilon_r = 90$ (Fig.~\ref{fig:fig2}). 

To ensure a controlled cross-platform comparison, the following modeling choices were harmonized across solvers: 
(i) coil geometry and dimensions; 
(ii) phantom geometry, dielectric properties, and placement; 
(iii) excitation configuration (quadrature drive and port locations); 
(iv) operating frequency; and 
(v) field normalization and post-processing procedures (details below).

\subsubsection{FDTD Implementation (Sim4Life)}
Simulations in Sim4Life
V8.0 employed an FDTD solver. A 3~mm coil mesh was selected following a convergence analysis across four resolutions (2-10~mm; Supplementary Methods), yielding a final discretization of 40 million cells; the 2~mm and 3~mm meshes produced identical primary resonant frequencies (63.56~MHz; $<0.001\%$ difference), whereas coarser meshes showed larger frequency deviations. Simulations used a broadband time step of $1.78\times10^{-10}$~s and were run for 250 RF periods until the $-30$~dB return-loss convergence criterion was satisfied. Absorbing boundaries were implemented using uniaxial perfectly matched layers (UPML) with a padding margin of 10~cm beyond the outermost coil structure.

\subsubsection{FEM Implementation (ANSYS HFSS)}
Simulations in ANSYS~HFSS~2021 employed adaptive tetrahedral mesh refinement. A length-based mesh operation was applied to the coil conductors, restricting the maximum surface element edge length to 10~mm. A length-based mesh operation (maximum element edge length 30~mm) was additionally applied inside the phantom volume. For the phantom, the final mesh contained approximately 630{,}000 tetrahedral elements. Convergence was monitored using the maximum change in $S$-parameters between consecutive adaptive passes ($\Delta S$), following our previously established framework \cite{Nguyen2020}. Simulations were considered converged when the maximum $\Delta S$ fell below 0.01 for phantom simulations and 0.002 for body models.

Final mesh statistics (element counts, edge lengths) and radiation-boundary padding are reported in Supplementary Methods.

\subsubsection{Field Normalization, Export, and Harmonized Post-processing}
\begin{figure*}[!t]
\centering
\includegraphics[width=\textwidth]{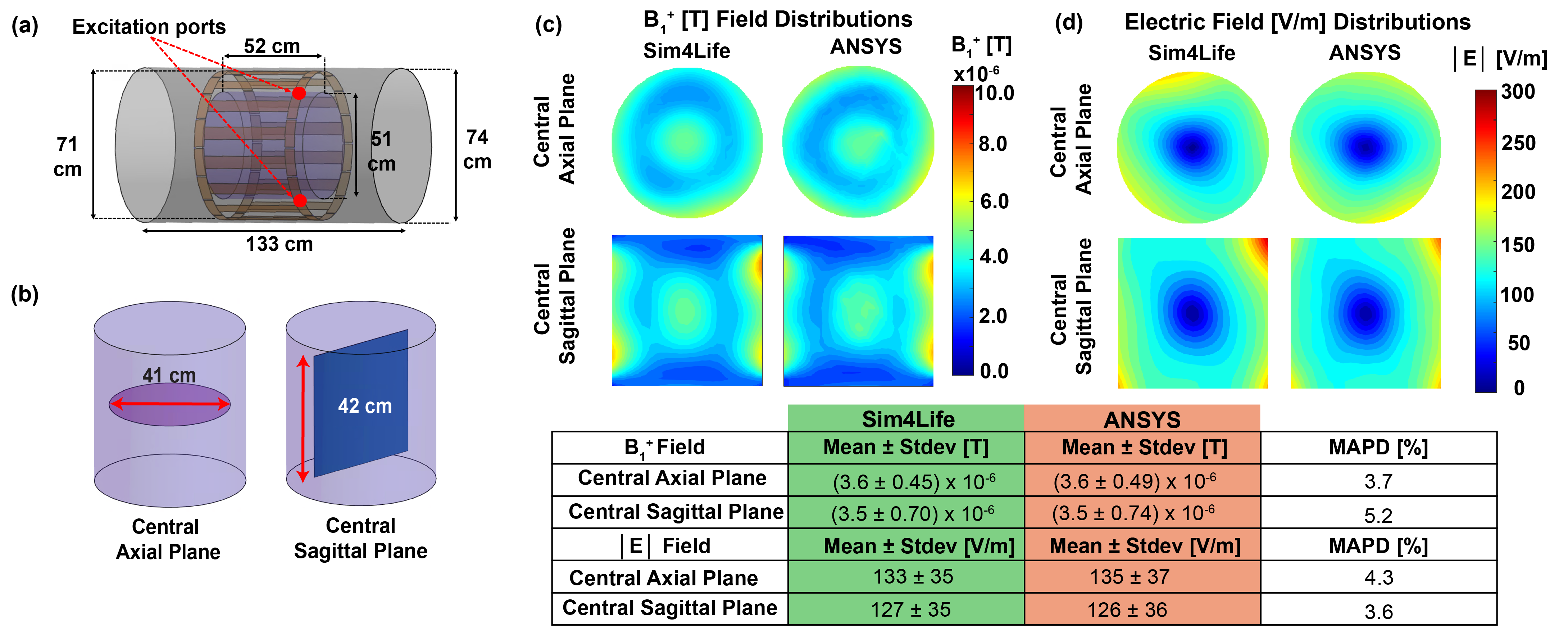}
\caption{Cross-platform comparison of simulated electromagnetic fields in a homogeneous phantom at 64~MHz (1.5~T). 
(a) Configuration of the 16-rung high-pass birdcage body coil loaded with a cylindrical phantom, detailing dimensions and excitation port placement. 
(b) Central axial and sagittal evaluation planes within the phantom. 
Simulated (c) $B_1^+$ field [T] and (d) electric field magnitude ($|E|$) [V/m] distributions generated by FDTD (Sim4Life) and FEM (ANSYS HFSS) solvers demonstrate excellent agreement. 
The summary table provides a quantitative comparison of the spatial mean, standard deviation, and mean absolute percentage difference (MAPD) across both planes.}
\label{fig:fig2}
\end{figure*}

To enable direct cross-platform comparison, all simulations were normalized to produce a mean $B_1^+$ magnitude of 4.9~$\mu$T on a 5-cm diameter central axial plane passing through the phantom isocenter. The $B_1^+$ and electric field distributions were extracted from central axial and sagittal planes for visualization and analysis (Fig.~\ref{fig:fig2}).

For quantitative comparisons (e.g., mean absolute percentage difference, MAPD), complex $B_1^+$ and electric-field data from both solvers were resampled onto a common uniform $300\times 300$ Cartesian grid over matched axial and sagittal evaluation planes (field-of-view: $41~\text{cm}\times 41~\text{cm}$ axial; $41~\text{cm}\times 42~\text{cm}$ sagittal) using linear interpolation (\texttt{scatteredInterpolant}, no extrapolation). All MAPD calculations were performed within a phantom-only mask to avoid boundary artifacts. $B_1^+$ was computed using the standard circularly polarized definition $B_1^+=(B_x+iB_y)/2$, and $|E|$ was computed as the magnitude of the complex electric field.

\subsection{AIMD Model and Rationale}

A commercially available deep brain stimulation (DBS) system (Vercise Cartesia\texttrademark, Boston Scientific) was selected as a representative elongated-lead AIMD to instantiate the ISO/TS~10974 Clause~8 Tier~3 workflow. DBS was chosen deliberately because its routing spans a large anatomical extent (intracranial lead to pectoral implantable pulse generator), thereby sampling heterogeneous loading conditions relevant to many elongated-lead AIMDs, and because it supports two clinically relevant but electromagnetically distinct configurations that provide a stringent test of RF coupling mechanisms.

Two device configurations were evaluated (Fig.~\ref{fig:fig3}B). In the lead-only configuration, a 45~cm lead (model DB-2202-45) was electrically capped at the proximal end and confined to the head region. In the full-system configuration, the same lead was connected to a 55~cm extension (model NM-3138-55) and an implantable pulse generator (IPG; model DB-1200-S), creating an approximately 100~cm conductive path traversing the head, neck, and chest. The difference in conductive pathway length and geometry between lead-only and full-system configurations creates distinct RF coupling pathways and, consequently, distinct heating scenarios.

While both configurations were used for transfer-function (TF) measurement and experimental validation (Section~\ref{sec:tf_validation}), subsequent \textit{in vivo} phenotype and surrogate analyses focused on the full-system configuration to capture the combined effects of whole-body loading and extended conductor routing under anatomically realistic conditions. Importantly, DBS is used here as an exemplar elongated-lead AIMD; the solver-equivalence methodology and conclusions are expected to extend to other elongated-lead AIMDs evaluated using ISO/TS~10974 Clause~8 Tier~3 workflows.

\subsection{Transfer Function Measurement and Experimental Validation} \label{sec:tf_validation}
\subsubsection{Tier 3 Transfer-Function Framework}
The Tier~3 transfer-function (TF) framework relates the complex tangential incident electric field along an implant trajectory to RF-induced heating at a target location. Following ISO/TS~10974 Clause~8 and prior work \cite{ISO10974,Park2007}, the RF-induced temperature rise due to an AIMD is computed as
\begin{equation}
\Delta T = C \left| \int TF(x)\,E_{\mathrm{tan}}(x)\,dx \right|^2 ,
\label{eq:deltaT}
\end{equation}
where $TF(x)$ is the complex-valued transfer function of the device, $E_{\mathrm{tan}}(x)$ is the complex tangential component of the incident electric field sampled along the device trajectory, and $C$ is a calibration factor.

In both solvers, the incident electric field was exported as complex phasor components at 63.6~MHz and sampled along the device centerline. The centerline was parameterized as a spline (Sim4Life, via the IMSAFE trajectory tool) and a polyline (ANSYS HFSS), and sampled at 1.0~mm intervals. The tangential electric field was defined as $E_{\mathrm{tan}}(x)=\vec{E}(x)\cdot \hat{t}(x)$. In Sim4Life, the unit tangent vector $\hat{t}(x)$ was obtained from the spline derivative. In ANSYS HFSS, the tangential component was computed directly using the Fields Calculator line tangent operator, i.e., $\mathrm{Dot}(\vec{E}, \mathrm{LineTangent})$, evaluated along the polyline, which is mathematically equivalent to the projection onto $\hat{t}(x)$. The integral in Eq.~\eqref{eq:deltaT} was evaluated over the lead-only ($0$ to $45$~cm) and full system conductive pathway ($0$ to $100$~cm).

The calibration factor was computed as
\begin{equation}
C = \frac{\Delta T_{\mathrm{measured}}}{\left| \int TF(x)\,E_{\mathrm{tan}}(x)\,dx \right|^2},
\label{eq:C}
\end{equation}
where $\Delta T_{\mathrm{measured}}$ is the measured temperature increase for a chosen reference configuration and exposure condition.
Calibration factors were derived independently for each solver and for each device configuration (lead-only and full system) using a single calibration trajectory selected to produce a mid-range temperature rise, and were then applied unchanged to predict $\Delta T$ across all remaining validation trajectories.

\subsubsection{Transfer-Function Measurement} 
Transfer functions for the evaluated DBS system were experimentally measured using a reciprocity-based approach \cite{Feng2015}. Measurements were conducted in a saline tank ($\sigma = 0.50$~S/m) using a vector network analyzer (E5063A, Keysight Technologies, Santa Rosa, CA, USA) to characterize the device’s RF coupling response along its length, as described in prior work \cite{Missoffe2018,Bhusal2025LowField,Jiang2023MRM}. The TF was reported as a complex-valued function of position along the lead/extension (Supplementary Fig.~S2).

\subsubsection{MRI Heating Experiments and Cross-Solver Prediction} 
RF-induced heating experiments were performed in a 1.5~T MRI system using an ASTM-compliant acrylic phantom filled with polyacrylic acid (PAA) gel ($\sigma = 0.50$~S/m, $\varepsilon_r=88$) (Supplementary Fig.~S1). A fiber-optic temperature probe (OSENSA, Vancouver, BC, Canada; resolution 0.01~$^\circ$C) was secured to the lead tip to measure temperature rise across distinct trajectories for lead-only ($n=15$) and full-system ($n=15$) configurations, generating a wide range of RF heating values (Supplementary Fig.~S1c). The phantom was positioned head-first supine, and experiments were performed at the chest landmark. A high-SAR T1-TSE sequence (acquisition time = 261~s, TE = 7.3~ms, TR = 836~ms, $B_1^+ = 4.9~\mu$T, FA = 180$^\circ$) was used for measurements. 

Computational incident electric fields extracted from each solver were coupled with the measured TFs to predict temperature rise via Eq.~\eqref{eq:deltaT}, and predictions were compared against experimental measurements for both device configurations. 

Agreement between predicted and measured temperature rise was quantified using Pearson correlation ($R$) and root-mean-square error (RMSE), reported separately for the lead-only and full-system configurations. Cross-solver equivalence of derived safety endpoints was assessed using two one-sided tests (TOST) as described in Section~\ref{sec:stats}.

\subsubsection{Experimental Uncertainty of Heating Validation Measurements}
\label{sec:exp_uncertainty}
Because the same TF measurements and the same experimental $\Delta T$ dataset were used for both solvers, experimental uncertainty does not differentially affect comparisons between HFSS and Sim4Life. However, quantifying the experimental uncertainty is important for interpreting absolute agreement between TF-based predictions and measured temperature rise (e.g., RMSE) and for selecting practically meaningful equivalence margins for $\Delta T$-based endpoints.

We estimated the experimental uncertainty of $\Delta T$ from run-to-run repeatability of RF exposure measurements and TF measurement acquisitions. Combining RF-exposure repeatability ($\sigma_\mathrm{exp} = 1.26\,^\circ$C) and TF measurement uncertainty ($\sigma_\mathrm{TF} = 0.22\,^\circ$C) in quadrature gave $\sigma_\mathrm{combined} = 1.28\,^\circ$C (Supplementary Methods, Fig.~S3). We defined an absolute equivalence margin $\Delta T_\mathrm{eq} = k\,\sigma_\mathrm{combined} = \pm 2.56\,^\circ$C ($k = 2$, specified a priori, corresponding to 95\% coverage) and used this as the margin for $p_{95}(\Delta T_\mathrm{ref})$ in TOST analyses. Probe-offset sensitivity (controlled offsets of $\pm 1$~mm and $\pm 2$~mm at ID4) produced a placement-induced $\Delta T$ range of $16.2\,^\circ$C ($9.2$--$25.4\,^\circ$C), reported separately as a sensitivity analysis (Supplementary Methods).

\begin{figure*}[!t]
\centering
\includegraphics[width=\textwidth]{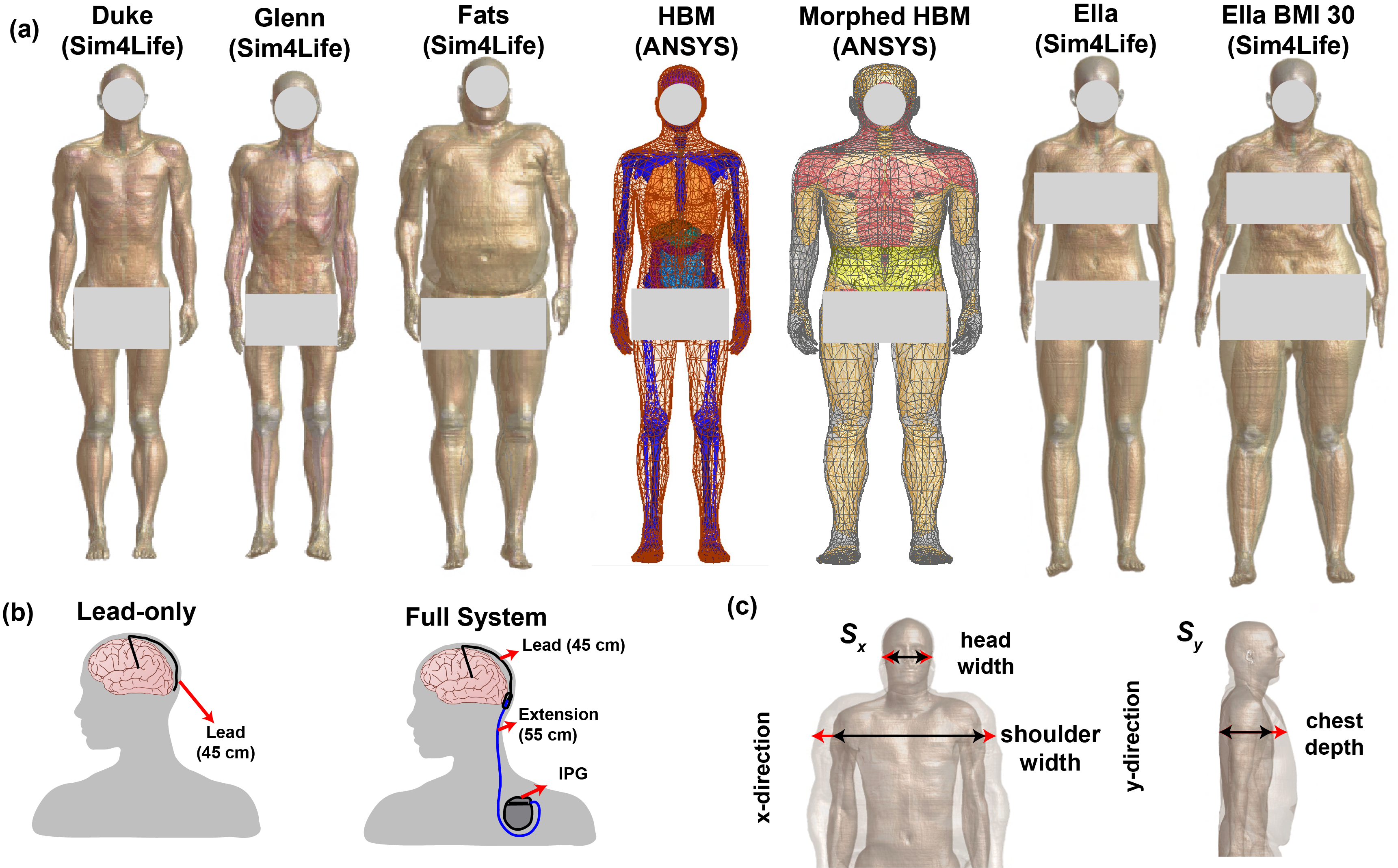}

\caption{Evaluated device configurations and anatomical models for \textit{in vivo} safety assessments.
(a) High-resolution computational anatomical models. Voxel-based models evaluated in Sim4Life include the standard adult male (Duke), an elderly male (Glenn), an obese male (Fats), a standard adult female (Ella), and a high-BMI female (Ella BMI 30). Surface-based models evaluated in ANSYS HFSS include the standard Human Body Model (HBM) and the geometrically morphed HBM approximating an obese phenotype.
(b) Deep brain stimulation (DBS) routing trajectories evaluated: a lead-only configuration (45~cm) and a full-system configuration (100~cm total; 45~cm lead, 55~cm extension, and IPG).
(c) Illustration of the geometric morphing strategy, detailing the scaling transformations applied to the standard HBM along the $x$-axis (shoulder width) and $y$-axis (chest depth) to approximate the volume of the obese phenotype.}
\label{fig:fig3}
\end{figure*}
\subsection{In Vivo Anatomical Modeling}
\label{sec:in_vivo_models}
\subsubsection{Anatomical Models and Positioning}
\textit{In vivo} heating assessments were conducted using high-resolution computational anatomical models spanning standard and diverse phenotypes and including both male and female anatomy. In Sim4Life, we evaluated voxel-based models from the Virtual Population~3.0 library, including Duke (standard adult male reference), Glenn (elderly male phenotype), Fats (obese male phenotype), Ella (standard adult female reference), and Ella BMI 30 (a high-BMI female model) \cite{Gosselin2014}. In ANSYS HFSS, we evaluated surface-based finite-element body models including the standard Human Body Model (HBM; male reference) and a geometrically morphed version of the male HBM (Morphed-HBM) used for surrogate analyses (Fig.~\ref{fig:fig3}). As shown in Table~\ref{tab:anatomical_models}, the anatomical models span a range of body types and representations.

All models were positioned head-first supine in the coil with the spine aligned and the head landmark at isocenter. Tissue dielectric properties were assigned using the IT'IS database. For Duke and HBM, dielectric properties were harmonized across platforms by mapping tissue classes and permittivity/conductivity values to an equivalent set. 

\subsubsection{Solver-Specific Discretization and Convergence}
For Sim4Life, a default voxel resolution of 2~mm was applied across all voxel models, with simulations converged to a return-loss threshold of $-30$~dB. To reduce computational load, Sim4Life body models were truncated at the torso level. The validity of this approximation was confirmed by comparing the complex tangential electric field ($E_{\mathrm{tan}}$) and predicted temperature rise along five representative full-system DBS trajectories between full-body and truncated Duke models, yielding complex $E_{\mathrm{tan}}$ correlations of 0.99 and a mean absolute difference in predicted $\Delta T$ of 0.01~$^\circ$C (RMSE = 0.009~$^\circ$C), indicating  negligible impact on transfer function predictions. For ANSYS HFSS, adaptive tetrahedral mesh refinement was employed with initial mesh operations of 20~mm for the body and 10~mm for the brain, with convergence criteria as described in Section~\ref{sec:rf_coil_modeling}. The final mesh consisted of approximately 60 million cells in Sim4Life and approximately 3 million elements in ANSYS HFSS for \textit{in vivo} simulations.
\subsubsection{Trajectory Generation and Harmonization Across Models}
Clinically realistic lead-only and full-system trajectories were generated based on post-operative patient CT data (approved by the institutional review board) including ipsilateral and contralateral orientations, and adapted to each anatomical model using CAD software (Rhino 7.0). Trajectory adaptation preserved key geometric features to the extent permitted by each model's anatomy, including cranial entry location, intracranial curvature, extracranial routing, and the IPG pocket location. Minor adjustments were permitted only to ensure anatomically valid paths.

For cross-platform equivalence analyses in standard male reference anatomies, we evaluated 50 lead-only and 50 full-system trajectories in Duke (Sim4Life) and the male HBM (HFSS). To assess sex-specific effects within Sim4Life, we evaluated full-system trajectories adapted to each of Duke and Ella (standard female reference), enabling a direct sex-stratified comparison at matched anatomical sites. An additional 50 full-system trajectories were evaluated in Ella BMI 30 to characterize the female obesity-driven risk profile. For phenotype comparisons in Sim4Life (Glenn and Fats) and ANSYS HFSS (Morphed HBM), we evaluated 50 full-system trajectories per model; 50 trajectories per model were used for the HFSS dielectric parameter sensitivity analyses.

\subsubsection{Sex-Stratified Analysis Within Sim4Life}
To quantify the impact of sex-specific anatomy on RF heating metrics, we performed a sex-stratified comparison between Duke (standard adult male) and Ella (standard adult female) within Sim4Life, using 50 matched full-system trajectories adapted to each model. For each model, we computed the distribution of $\Delta T$ across trajectories, the 95th-percentile temperature rise $p_{95}$ at the reference exposure, and the derived Maximum Allowable $B_1^+$ limit. TOST equivalence testing (Section~\ref{sec:stats}) was additionally applied to assess whether the two standard-BMI models yield equivalent imaging-relevant endpoints. To characterize the female obesity risk profile, we further compared Ella against Ella BMI 30 using the same trajectory set and endpoints.


\subsection{Evaluation of Risk Approximation Strategies}
\label{sec:surrogates}
To approximate the elevated RF-heating risk observed in the obese voxel model (Fats) using standard, readily available anatomical inputs, we evaluated two surrogate strategies that can be implemented within common commercial workflows: (A) tissue dielectric property variation and (B) geometric morphing. Unless otherwise stated, surrogate analyses were performed using the ANSYS HFSS male HBM as the reference model, enabling direct assessment of surrogate strategies in surface-based FEM models.

\subsubsection{Strategy A: Dielectric Parameter Sensitivity (Tissue Uncertainty)}
A sensitivity analysis was performed to assess whether plausible uncertainty in tissue dielectric properties could reproduce the elevated heating tail observed in the obese phenotype. In the HFSS male HBM, both conductivity ($\sigma$) and relative permittivity ($\varepsilon_r$) were scaled together for key tissues encountered along the implant path: blood, brain, cerebellum, muscle, fat, and skin. Four uniform scaling conditions were evaluated ($-50\%$, $-25\%$, $+25\%$, $+50\%$ from nominal values), along with two random-scaling conditions in which each tissue received an independent scaling factor uniformly sampled from 0.50 to 1.50 (Supplementary Table~S1). 

For each scaling condition, $\Delta T$ distributions were computed across the full-system trajectory set, and the ``worst-case sweep'' was defined as the condition producing the maximum $p_{95}(\Delta T)$ at the reference exposure. 

\subsubsection{Strategy B: Geometric Morphing (Anatomical Approximation)}

The standard HFSS male HBM was geometrically morphed to approximate the macroscopic external dimensions of the obese phenotype observed in voxel models. Scaling factors were derived from anatomical dimension ratios between the obese (Fats) and standard (Duke) voxel models. Specifically, the $x$-axis (lateral, $S_x$) scaling factor was computed as the average of head width (174/152~mm = 1.145) and shoulder width (582/483~mm = 1.204), yielding $S_x = 1.175$, and the $y$-axis (anterior--posterior, $S_y$) scaling factor was derived from chest depth (265/210~mm = 1.26), yielding $S_y = 1.26$ (Fig.~\ref{fig:fig3}C). No scaling was applied along the $z$-axis ($S_z = 1.0$). This transformation was applied to the entire body volume, with internal organs scaling proportionally.


\subsubsection{Homogenized Verification (Optional Low-Information Surrogate)}

To assess whether simplified tissue modeling could approximate worst-case predictions, we evaluated a homogenized version of the morphed model in which a single effective dielectric property was assigned throughout the body and varied by $\pm50\%$ around its nominal value. This analysis was performed for the Morphed HBM and compared against the heterogeneous Morphed HBM and the native obese-phenotype target (Fats).

\subsubsection{Evaluation Endpoints and Comparison to Obese-Phenotype Target}

Surrogate strategies were evaluated using the same imaging-relevant endpoints used throughout the study: (i) the distribution of predicted $\Delta T$ across trajectories at the reference exposure, (ii) the 95th-percentile temperature rise $p_{95}(\Delta T)$, and (iii) the derived Maximum Allowable $B_1^+$ limit required to satisfy the $p_{95}(\Delta T)\le 2^\circ$C criterion (Section~\ref{sec:thresholds}). Performance was assessed by comparing each surrogate's $p_{95}$ and Maximum Allowable $B_1^+$ to those obtained in the native obese phenotype model (Fats) and by evaluating whether the surrogate correctly identifies the direction and magnitude of risk elevation relative to standard reference models. 

\subsection{Clinical Threshold Analysis}
\label{sec:thresholds}

To translate predicted heating distributions into scanner-operational safety constraints, we defined a model-specific ``Maximum Allowable $B_1^+$'' for each anatomical model and implant configuration. This metric was selected as the primary safety constraint because current MR Conditional labeling practice emphasizes scanner-controllable RF exposure metrics (e.g., $B_{1,\mathrm{rms}}^+$) and because $B_1^+$ provides a consistent normalization target across simulated coil models \cite{ISO10974}. For each model, the Maximum Allowable $B_1^+$ was computed such that the 95th percentile of the predicted temperature-rise distribution across trajectories did not exceed a conservative threshold of $2^\circ$C:
\begin{equation}
p_{95}\!\left(\Delta T(B_1^+)\right) \le 2^\circ\mathrm{C}.
\label{eq:p95criterion}
\end{equation}

Because $\Delta T$ is proportional to deposited RF power, and deposited power scales with the square of the incident electric field, $\Delta T$ scales quadratically with $B_1^+$. Therefore, scaling from the reference simulation at $B_{1,\mathrm{ref}}^+$ to a candidate exposure $B_{1,\mathrm{new}}^+$ was governed by
\begin{equation}
\Delta T_{\mathrm{new}} = \Delta T_{\mathrm{ref}}\left(\frac{B_{1,\mathrm{new}}^+}{B_{1,\mathrm{ref}}^+}\right)^2 .
\label{eq:scaling}
\end{equation}
For each anatomical model (including male and female reference models) and each device configuration analyzed, the Maximum Allowable $B_1^+$ was computed by applying Eq.~\eqref{eq:scaling} to the trajectory-wise $\Delta T_{\mathrm{ref}}$ distribution and solving for the largest $B_{1,\mathrm{new}}^+$ that satisfies Eq.~\eqref{eq:p95criterion}.

Unless otherwise stated, reference distributions were generated at $B_{1,\mathrm{ref}}^+=4.9~\mu$T after 261~s RF exposure (consistent with the high-SAR T1-TSE validation sequence), and all reported $p_{95}$ values correspond to this reference exposure prior to scaling.

\subsubsection{Uncertainty Estimation for Percentile-Based Endpoints}

Confidence intervals (CIs) for $p_{95}(\Delta T_{\mathrm{ref}})$ and for the derived Maximum Allowable $B_1^+$ were estimated using nonparametric bootstrap resampling of trajectories. Specifically, 10{,}000 bootstrap samples were generated by resampling trajectories with replacement, and for each sample we computed $p_{95}(\Delta T_{\mathrm{ref}})$ and the corresponding Maximum Allowable $B_1^+$ using Eq.~\eqref{eq:scaling}. The reported 95\% CIs correspond to the 2.5 and 97.5 percentiles of the bootstrap distribution.
\begin{figure}[!b]
\centering
\includegraphics[width=\columnwidth]{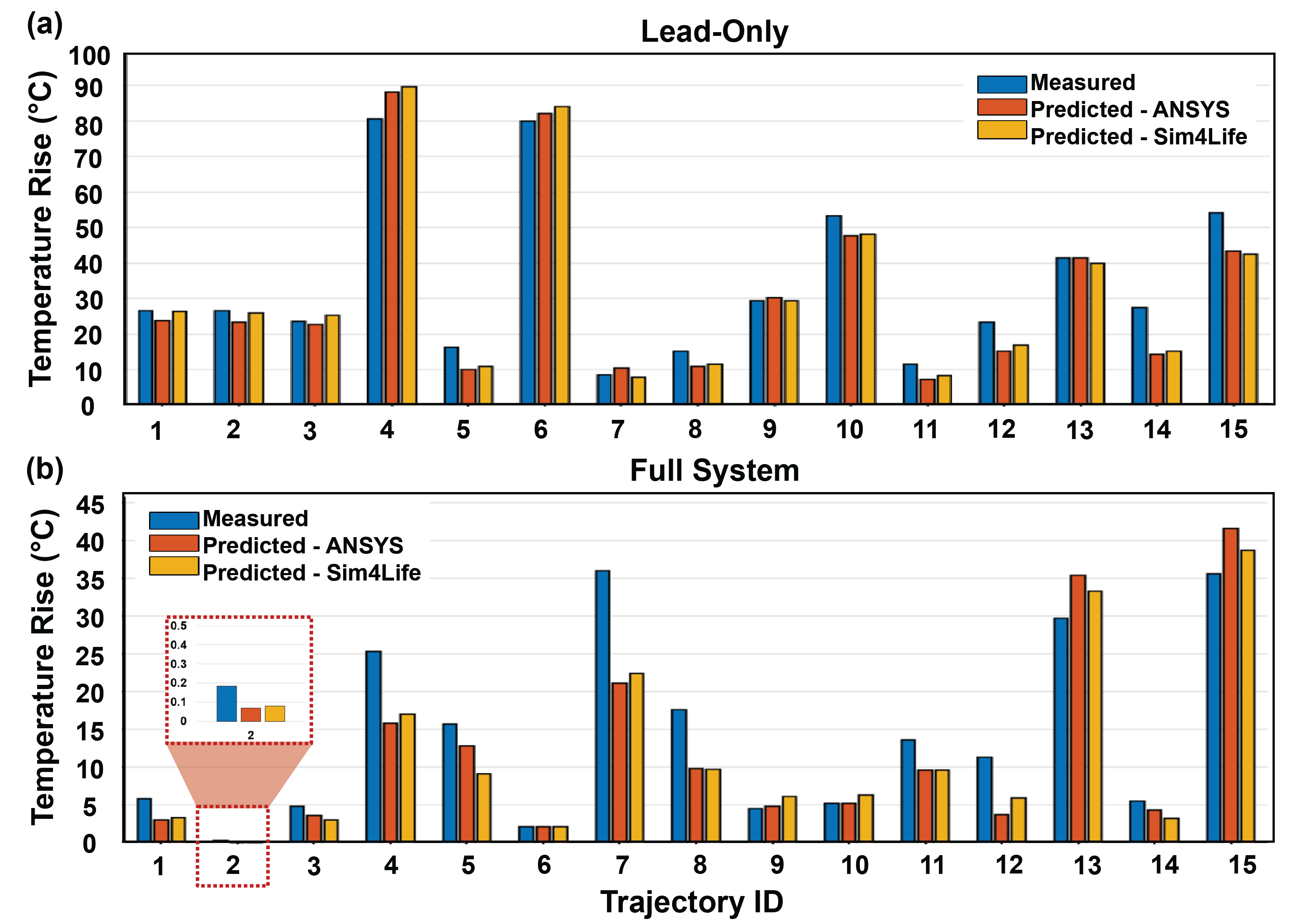}
\caption{Transfer function measurements and validation of RF-induced heating. Comparison of measured and computationally predicted temperature rises ($\Delta T$) across 15 distinct lead-only (a) and full system (b) trajectories. Predictions derived from both FDTD (Sim4Life) and FEM (ANSYS HFSS) incident fields demonstrate strong agreement with physical measurements for both the lead-only ($R=0.98$ for both solvers) and full-system ($R \ge 0.90$) configurations.}
\label{fig:fig4}
\end{figure}
\subsubsection{Rationale for Thermal Threshold}

A $2^\circ$C threshold was adopted as a conservative safety margin for deep brain stimulation, consistent with prior DBS MRI safety literature and thermal dose considerations \cite{Boutet2020}, and was applied uniformly across all models and trajectories.

\subsection{Agreement, Equivalence, and Statistical Analysis}
\label{sec:stats}

\subsubsection{Agreement Metrics}
Cross-solver agreement for phantom field distributions (Section~\ref{sec:rf_coil_modeling}) was quantified using the mean absolute percentage difference (MAPD) computed within a phantom-only mask. Agreement between TF-based predictions and measured temperature rise in the heating validation experiments (Section~\ref{sec:tf_validation}) was quantified using Pearson correlation ($R$) and root-mean-square error (RMSE). Cross-solver agreement in TF-based $\Delta T$ predictions across matched trajectories was additionally quantified using Bland--Altman analysis (bias and 95\% limits of agreement, LoA), with log-transformed analysis applied to account for heteroscedasticity (Supplementary Fig.~S4). For \textit{in vivo} simulations, trajectory-wise $\Delta T$ distributions were summarized using the median, interquartile range, and the 95th percentile $p_{95}(\Delta T_{\mathrm{ref}})$ at the reference exposure, along with the derived Maximum Allowable $B_1^+$ (Section~\ref{sec:thresholds}).

\subsubsection{Equivalence Testing Using Two One-Sided Tests (TOST)}
Cross-solver comparisons were treated as an equivalence problem rather than a null-hypothesis ``difference'' test, applying two one-sided tests (TOST) at $\alpha = 0.05$. The primary equivalence endpoint was $p_{95}(\Delta T_{\mathrm{ref}})$ at the reference exposure, with an absolute equivalence margin of $\pm\Delta T_{\mathrm{eq}}$ derived \emph{a priori} from the experimentally estimated uncertainty of $\Delta T$ (see Section~\ref{sec:exp_uncertainty} and Supplementary Methods). Equivalence was concluded when the 95\% CI of the bootstrap-derived endpoint difference lay entirely within this margin. Maximum Allowable $B_1^+$ limits were reported as point estimates with 95\% bootstrap  confidence intervals; for reference, the uncertainty-equivalent relative margin  for $B_1^+$ is $\pm 13.1\%$, derived by mapping $\Delta T_{\mathrm{eq}} = \pm 2.56~^\circ$C through the quadratic $B_1^+$-$\Delta T$ scaling relationship at the safety threshold (see Supplementary Methods).

\subsubsection{Secondary Nonparametric Tests}
Because trajectory-wise $\Delta T$ distributions were non-Gaussian, secondary analyses comparing distribution shapes across phenotypes or surrogate strategies used nonparametric tests. Specifically, the unpaired Wilcoxon rank-sum test was used for independent samples, with significance assessed at $p<0.05$ (no multiple-comparison correction was applied). Surrogate strategies were evaluated primarily using the percentile-based endpoints ($p_{95}$ and Maximum Allowable $B_1^+$) and their bootstrap confidence intervals.

\subsubsection{Software}
All statistical analyses were performed using MATLAB R2023b (MathWorks, Natick, MA) for TOST, Bland--Altman analysis, and bootstrap resampling.

\begin{figure}[!t]
\centering
\includegraphics[width=\columnwidth]{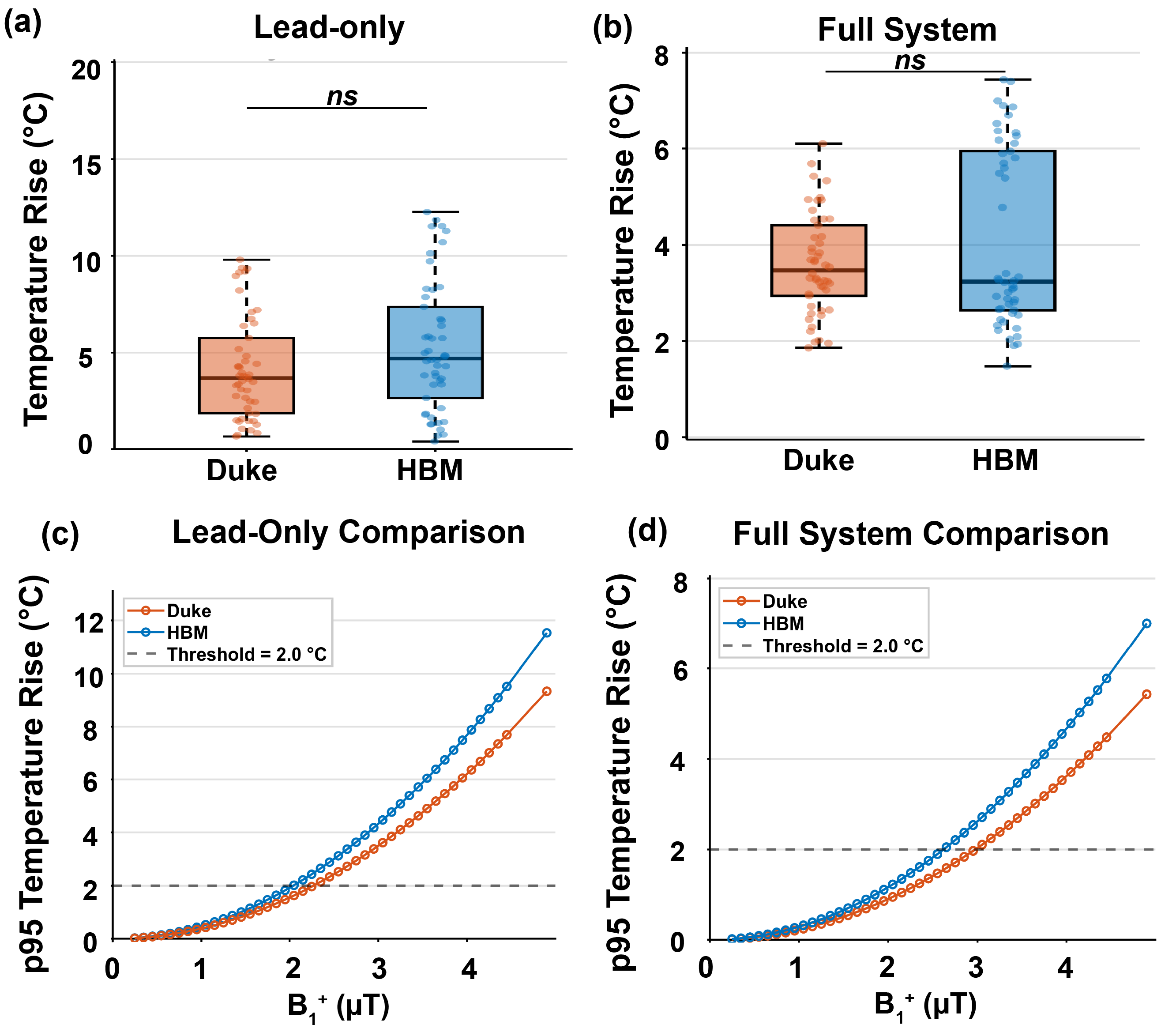}
\caption{Cross-platform \textit{in vivo} equivalence between standard reference models. Boxplots comparing the distribution of predicted temperature rises ($\Delta T$) across routing trajectories for the (a) lead-only and (b) full-system configurations using the voxel-based Duke model (FDTD/Sim4Life) and the surface-based HBM (FEM/ANSYS HFSS). Both configurations showed no statistically significant difference (n.s., $p>0.05$). The derived Maximum Allowable $B_1^+$ safety limits are shown for the (c) lead-only and (d) full-system configuration.}
\label{fig:fig5}
\end{figure}

\section{Results}
\subsection{Cross-Platform Field Equivalence}
Electromagnetic field distributions generated by the FDTD (Sim4Life) and FEM (ANSYS HFSS) solvers showed strong agreement in the homogeneous phantom at 64~MHz (Fig.~\ref{fig:fig2}). After normalization to a mean $B_1^+$ magnitude of 4.9~$\mu$T on the central 5-cm-diameter axial plane, the mean absolute percentage difference (MAPD) in $B_1^+$ magnitude was 3.7\% in the central axial plane and 5.2\% in the sagittal plane. Electric-field magnitude ($|E|$) comparisons yielded MAPD values of 4.3\% (axial) and 3.6\% (sagittal). Across both planes and both field quantities, MAPD values remained below 6\%, supporting close agreement of incident-field calculations between the two solver paradigms under a controlled, like-for-like setup.

\subsection{Experimental Validation of Transfer Function Predictions}
Transfer function (TF)--based temperature rise predictions demonstrated strong agreement with experimental measurements across both device configurations (Fig.~\ref{fig:fig4}). For the lead-only configuration, predicted temperature rises were highly correlated with measurements ($R=0.98$ for both solvers), with RMSE of 5.86~$^\circ$C (Sim4Life) and 6.0~$^\circ$C (HFSS). For the full-system configuration, correlations remained strong ($R\ge 0.90$), with RMSE of 5.5~$^\circ$C (Sim4Life) and 5.9~$^\circ$C (HFSS). Bland--Altman analysis of solver-to-solver agreement across matched trajectories yielded a mean bias of $1.1~^\circ$C (LoA: [$-3.2$, $+5.3$]~$^\circ$C) for the lead-only configuration, with Sim4Life predictions slightly exceeding HFSS; the full-system configuration showed near-zero bias ($-0.20~^\circ$C, LoA: [$-3.4$, $+3.0$]~$^\circ$C). Log-transformed analyses confirmed low proportional bias across both configurations (Supplementary Fig.~S4).
Calibration factors ($C$) were derived independently for each solver and device configuration from a single mid-heating calibration trajectory and applied unchanged to all remaining validation trajectories. For the lead-only configuration, $C$ differed by 7.3\% between solvers. 
For the full-system configuration, $C$ differed by only 2.8\%.


\subsection{Cross-Platform In Vivo Equivalence in Standard Reference Models (Male)} 
Trajectory-wise $\Delta T$ distributions were compared between Duke (Sim4Life) and the male HBM (HFSS) for both configurations (Fig.~5); neither showed a statistically significant difference under rank-based testing ($p > 0.05$). At $B_1^{+} = 4.9~\mu$T, the lead-only configuration gave $p_{95} = 9.3\,^\circ$C (95\% CI: 8.2--9.8) in Duke and $11.5\,^\circ$C (10.1--12.3) in HBM, with Maximum Allowable $B_1^{+}$ limits of 2.27~$\mu$T (2.21--2.42) and 2.04~$\mu$T (1.98--2.18). The full-system configuration gave $p_{95} = 5.4\,^\circ$C (4.9--6.1) and $7.0\,^\circ$C (5.5--7.4), with limits of 2.97~$\mu$T (2.80--3.12) and 2.62~$\mu$T (2.54--2.71). TOST confirmed endpoint equivalence for $p_{95}(\Delta T_\mathrm{ref})$ (difference 95\% CI $[+0.77, +2.46]\,^\circ$C, within the $\pm 2.56\,^\circ$C margin); the 0.36~$\mu$T $B_1^{+}$ point difference ($-12\%$ relative to Duke) corresponds to only $0.58\,^\circ$C above the $2\,^\circ$C threshold, within the $\pm 13.1\%$ uncertainty-equivalent margin.

\subsection{Sex-Stratified Analysis: Duke vs. Ella (Sim4Life)}
To assess sex at matched BMI, Duke ($n = 50$) and Ella ($n = 50$ matched trajectories) were compared within Sim4Life. At $B_1^{+} = 4.9~\mu$T, $p_{95}$ was $5.4\,^\circ$C (4.9--6.1) and $6.6\,^\circ$C (6.2--7.4), with Maximum Allowable $B_1^{+}$ of 2.97~$\mu$T (2.80--3.12) and 2.68~$\mu$T (2.54--2.78); distributions did not differ ($p = 0.22$). TOST confirmed equivalence ($p_{95}$ difference 95\% CI $[+0.31, +2.1]\,^\circ$C, within $\pm 2.56\,^\circ$C); the $B_1^{+}$ difference was 0.29~$\mu$T ($-10.1\%$ relative to Duke).

\begin{figure}[!t]
\centering
\includegraphics[width=\columnwidth]{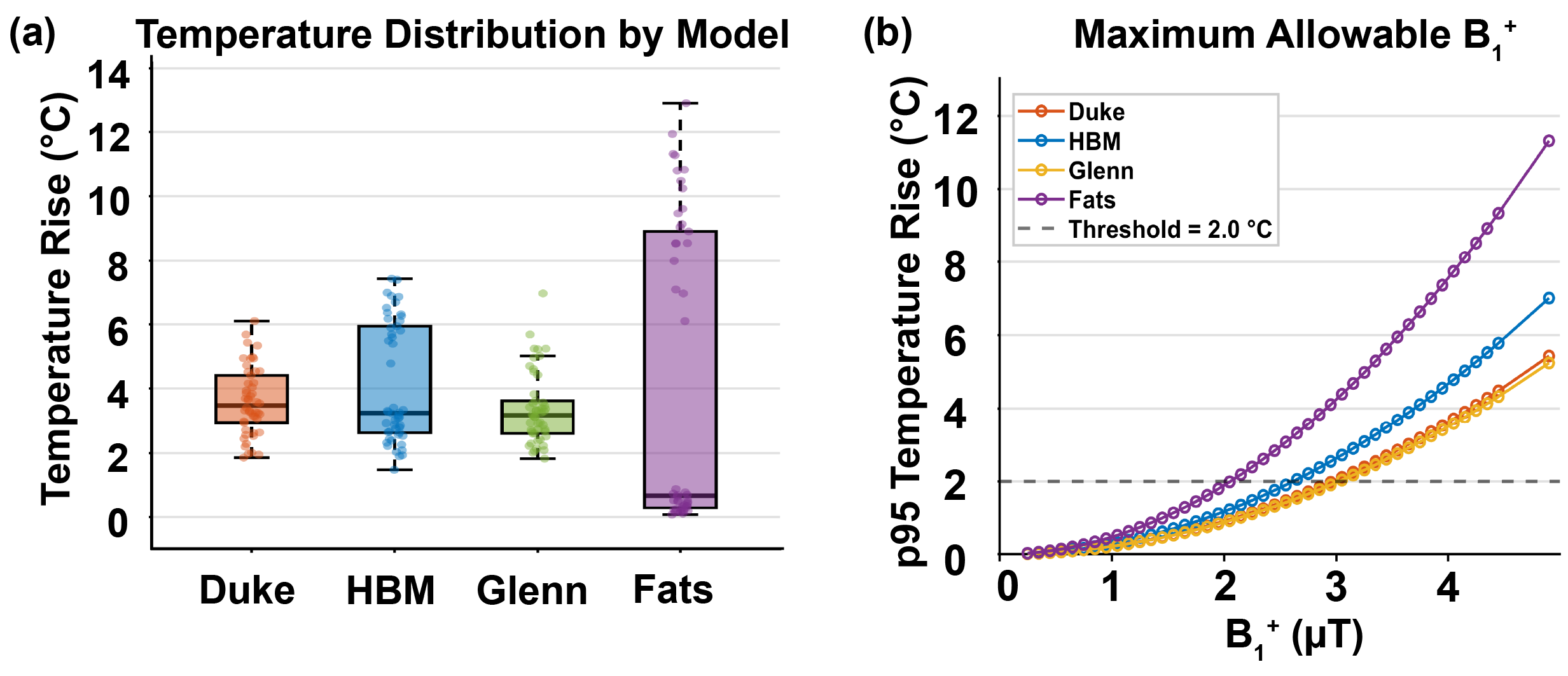}
\caption{
Impact of anthropometric variability on RF-induced heating. (a) Temperature rise distributions across standard (Duke, HBM), elderly (Glenn), and obese (Fats) anatomical models at the reference $B_1^+$ = 4.9 ~$\mu$T. (b) Derived Maximum Allowable $B_1^+$ limits based on a 2°C threshold. Standard and elderly models yielded comparable safety limits (2.62--3.03~$\mu$T).}
\label{fig:fig6}
\end{figure}

\subsection{Impact of Anthropometric Variability}
Standardized reference models produced convergent safety endpoints, whereas heating risk shifted substantially across diverse phenotypes (Fig.~\ref{fig:fig6}). The elderly model (Glenn) produced a $p_{95}$ of $5.2\,^\circ$C (95\% CI: 4.9--6.9) at $B_1^+ = 4.9~\mu$T, with no statistically significant difference from Duke for the full-system configuration ($p > 0.05$); its derived Maximum Allowable $B_1^+$ limit was 3.03~$\mu$T (95\% CI: 2.62--3.11), overlapping both Duke and HBM. In contrast, the obese model (Fats) showed a substantially higher tail ($p_{95} = 11.3\,^\circ$C, 95\% CI: 10.4--12.9), more than a two-fold increase in near-worst-case heating relative to standard models, reducing the Maximum Allowable $B_1^+$ limit to 2.06~$\mu$T (95\% CI: 1.93--2.15). This is consistent with prior observations that body composition and subcutaneous fat significantly alter RF coupling and power deposition around elongated implants \cite{BhusalPMC, Bhusal2021JMRI}.

\subsection{Female High-BMI: Ella vs.\ Ella BMI 30}
\label{sec:ellavella30}
To test whether the obesity-driven risk amplification observed in the male cohort (Fats vs.\ Duke) generalizes to female anatomy, we evaluated Ella BMI~30 using the same 50 full-system trajectories. Ella BMI~30 produced higher heating distribution: $p_{95} = 10.3~^\circ$C (95\%~CI: [9.6, 11.4]), representing a 55\% increase relative to standard-BMI Ella ($p_{95} = 6.6~^\circ$C) (Fig.~\ref{fig:fig7}C). The Maximum Allowable $B_1^+$ limit decreased to $2.16~\mu$T (95\%~CI: [2.05,~2.24]), a 19\% reduction compared to Ella ($2.68~\mu$T), and a reduction comparable to that observed for the obese male phenotype (Fats vs.\ Duke: 31\%). These results confirm that anatomy-driven BMI loading effects on RF heating are consistent across patient sex.

\subsection{Evaluation of Surrogate Strategies}
To determine practical methods for approximating the elevated risk observed in the obese model (Fats) using standard anatomical inputs, we evaluated two surrogate strategies: dielectric-property variation and geometric morphing (Fig.~\ref{fig:fig7}). Surrogates were judged by their ability to reproduce (i) the shift in $p_{95}(\Delta T_{\mathrm{ref}})$ and (ii) the corresponding reduction in Maximum Allowable $B_1^+$ relative to reference models.

\subsubsection{Dielectric Parameter Sensitivity}
A comprehensive dielectric-parameter sensitivity analysis of the standard HFSS HBM (uniform and random scaling up to $\pm50\%$) failed to reproduce the obese high-risk tail. Across all dielectric variations, $p_{95}$ ranged from 6.8$^\circ$C to 8.3$^\circ$C, yielding Maximum Allowable $B_1^+$ limits of 2.40--2.60~$\mu$T, which remained comparable to the baseline HBM and substantially above the obese-derived limit (Supplementary Fig.~S5).

\subsubsection{Geometric Morphing}
In contrast, the geometrically Morphed HBM model produced a $p_{95}$ of $11.3~^\circ$C (95\%~CI: [10.8,~12.1]) at $B_1^+ = 4.9~\mu$T, approximating the native obese tail of Fats ($p_{95} = 11.3~^\circ$C, 95\%~CI: [10.4,~12.9]), with Maximum Allowable $B_1^+$ limits of 2.06~$\mu$T (95\%~CI: [2.00,~2.11]) and 2.06~$\mu$T (95\%~CI: [1.93,~2.14]) respectively. Notably, the worst-case trajectory in both models belonged to the same routing category (a contralateral trajectory with two $\approx3~cm$ loops). This indicates that geometric morphing preserved the spatial RF coupling pattern responsible for worst-case heating, and that the surrogate correctly identifies the class of configurations driving peak risk.

To assess whether simplified tissue modeling could approximate worst-case predictions, a homogenized Morphed HBM was evaluated under a similar dielectric sweep ($\pm50\%$). The homogenized Morphed HBM yielded $p_{95}$ values of 7.7$^\circ$C to 9.9$^\circ$C (Maximum Allowable $B_1^+$: 2.20--2.50~$\mu$T), remaining below the heterogeneous Morphed HBM and below native Fats (Supplementary Fig.~S6).

\subsubsection{Female Robustness Check}
Cross-sex corroboration of the obesity-driven risk elevation is provided by the Sim4Life Ella BMI 30 results (Section~\ref{sec:ellavella30}): the female high-BMI model produced a Maximum Allowable $B_1^+$ of $2.16~\mu$T, consistent in direction and magnitude with the male Fats result ($2.06~\mu$T), supporting the generalizability of geometry-dominant loading effects across anatomical sex.

\subsection{Clinical Threshold Summary}
\begin{figure*}[!t]
\centering
\includegraphics[width=\textwidth]{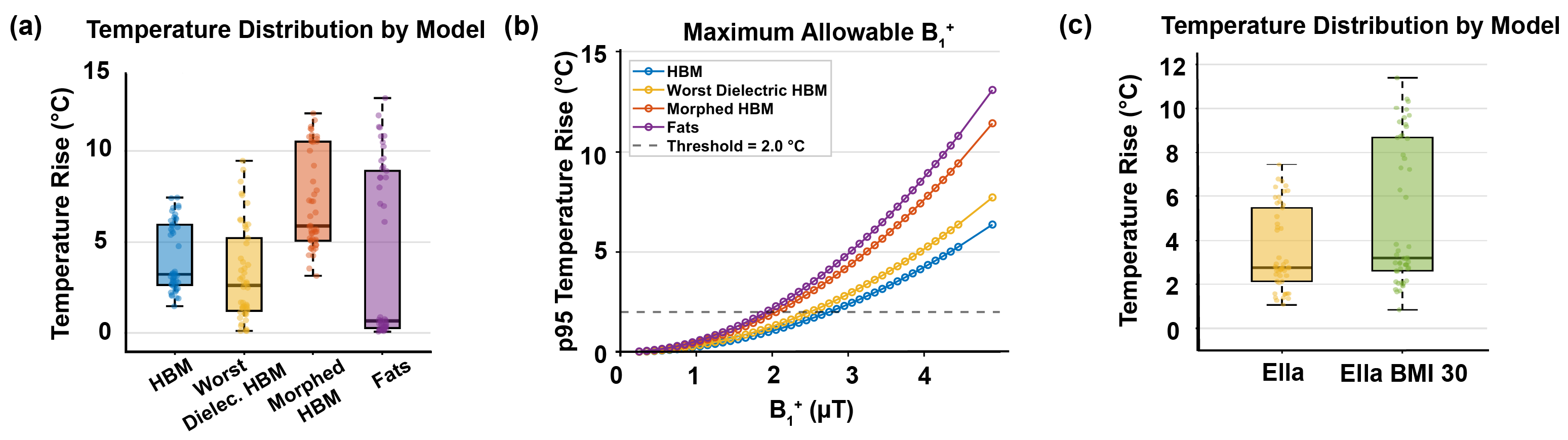}
\caption{Evaluation of surrogate strategies and cross-sex corroboration of obese-phenotype risk. (a) Temperature rise distributions for the standard HBM, worst-case dielectric sweep, geometrically Morphed HBM, and native obese model (Fats) at the reference $B_1^+$ = 4.9~$\mu$T. (b) Derived Maximum Allowable $B_1^+$ limits based on a 2$^\circ$C threshold; the dielectric sweep failed to shift heating beyond the standard HBM range, while geometric morphing approximated the Fats threshold. (c) Temperature rise distributions for standard-BMI (Ella) and elevated-BMI (Ella BMI~30) female models across 50 full-system trajectories, with $p_{95}$ increasing from $6.6$ to $10.3~^\circ$C and Maximum Allowable $B_1^+$ decreasing from $2.68$ to $2.16~\mu$T.}
\label{fig:fig7}
\end{figure*}

Derived Maximum Allowable $B_1^+$ constraints clustered near $\sim3~\mu$T for standard reference models (Duke, HBM, Glenn, Ella) but dropped to $\sim2.1~\mu$T for both obese phenotypes (Fats, Ella BMI 30; Fig.~\ref{fig:fig6}, Fig.~\ref{fig:fig7}). Geometric morphing reproduced the obese male constraint (Morphed HBM: 2.06~$\mu$T), whereas dielectric sweeps did not.

\section{Discussion}

\subsection{Solver Equivalence and the Dominance of Anatomy}
This study addresses a critical ambiguity in the regulatory assessment of AIMD safety: determining whether the choice of computational solver or the selection of anatomical model is the primary source of predictive uncertainty. Our cross-platform validation demonstrates that when simulation parameters and anatomical inputs are standardized, FDTD (Sim4Life) and FEM (ANSYS HFSS) solvers produce practically equivalent results. We observed excellent agreement in phantom field distributions (MAPD $< 6\%$) and strong correlations with experimental measurements for both lead-only and full-system configurations ($R \ge 0.90$). Furthermore, \textit{in vivo} predictions for standard body models (Duke vs.\ HBM) converged on practically indistinguishable safety thresholds, with $p_{95}$ differences falling within the pre-specified experimental uncertainty margin.

These findings suggest that the choice of numerical engine is not a significant driver of variability in Tier 3 assessments. Instead, our results identify patient anthropometry, specifically BMI, as the dominant source of variability. 

\subsection{The Hierarchy of Risk: Geometry Overrides Tissue Properties}
A central finding of this work is the dominance of geometry over tissue dielectric uncertainty. Current regulatory practices often emphasize dielectric sensitivity analyses to quantify uncertainty \cite{ISO10974}. However, our results demonstrate that such analyses, even with $\pm50\%$ variation, are insufficient to capture the heating disparity associated with elevated BMI, shifting the 95th-percentile by less than 2$^\circ$C. Geometrically morphing the body to approximate the obese phenotype shifted the distribution by $>4^\circ$C, indicating that anthropometric variation represents a distinct source of uncertainty not addressed by dielectric sweeps alone.

This implies that worst-case heating is driven principally by global electromagnetic coupling and coil-loading effects dictated by patient size, rather than local conductivity variations at the electrode tip. This observation aligns with recent work by Carluccio \textit{et al.}, who showed that anatomical complexity can often be simplified without compromising safety predictions provided the macroscopic field distribution is accurate \cite{Carluccio2021}, and Bottauscio \textit{et al.}, who highlighted patient anatomy as a primary determinant of heating variability \cite{Bottauscio2024}. Consequently, regulatory efforts might benefit from prioritizing the inclusion of anthropometrically diverse models (or geometric surrogates) over exhaustive dielectric sweeps of standard models.

\subsection{BMI as the Dominant Variable}
Age did not independently elevate heating risk (Duke vs.\ Glenn: $p > 0.05$), whereas BMI was the dominant driver of variability. The obese male model (Fats, BMI 36) required a 31\% reduction in safe exposure limits, indicating that macroscopic body geometry predicts RF heating more strongly than tissue dielectric variation for the configurations evaluated; only one elderly model was assessed, so age-matched comparisons remain to be extended across additional phenotypes.

This BMI-driven elevation held within female anatomy. At standard BMI, Duke and Ella produced comparable distributions ($p = 0.22$), with TOST confirming endpoint equivalence despite a 0.29~$\mu$T difference in Maximum Allowable $B_1^{+}$ (2.97 vs.\ 2.68~$\mu$T), whereas Ella BMI 30 showed an elevated profile. Fats represents a larger BMI increment than Ella BMI~30 (13.8 vs.\ 8.4 units above their respective references), which accounts for its larger absolute reduction. Normalizing by increment yields a consistent per-unit effect of approximately 2.2--2.3\% reduction in Maximum Allowable $B_1^+$ per BMI unit across both sexes, identifying BMI increment as the primary determinant of risk elevation over the range evaluated and supporting BMI-stratified exposure limits for elongated-lead AIMDs regardless of patient sex \cite{Yao2019, Alon2016}.

\subsection{Geometric Morphing as a Practical Surrogate}
Geometric morphing strategy provides an accessible way to approximate this risk. The geometrically Morphed HBM produced a Maximum Allowable $B_1^+$ of  2.06~$\mu$T, converging on the same point estimate as the native Fats model, with overlapping bootstrap confidence intervals ([2.00,~2.11] vs.\  [1.93,~2.14]~$\mu$T).

However, the Morphed HBM did not fully replicate the native obese distribution shape; the morphed model predicted higher average temperature rise across all trajectories despite converging on the same $p_{95}$ and Maximum Allowable $B_1^+$ endpoints. This suggests that simple geometric scaling is sufficient for approximating worst-case safety limits, even without explicitly modeling adipose tissue redistribution associated with obesity.

From a workflow perspective, morphing offers a practical method for estimating elevated BMI risk when native high-BMI models are unavailable, though the resulting limits should be interpreted as approximations rather than strict bounds.

\subsection{Limitations}
Several limitations of this study should be acknowledged. First, this study was conducted at 1.5~T; extension to 3~T is warranted given the different RF wavelength characteristics and potentially altered implant resonance behavior at higher field strengths. Furthermore, emerging low-field scanners (e.g., 0.55 T), field shaping methods, and open-bore vertical (e.g., 1.2 T) MRI systems introduce fundamentally different electric and magnetic field distributions \cite{Bhusal2025LowField, Sanpitak2025, Golestanirad2020, Kazemivalipour2020}. Specifically, hardware-driven field shaping techniques such as parallel radiofrequency transmission (pTx) and reconfigurable rotating coil arrays dynamically manipulate the amplitude and phase of the incident RF to steer electric-field nulls toward the implant trajectory, presenting entirely different boundary conditions than conventional uniform excitation \cite{McElcheran2015, McElcheran2017, Golestanirad2017, Kutscha2025}. Validating cross-platform solver equivalence in these highly modified and novel magnetic field environments is a necessary next step. Similarly, validation across other AIMD types (e.g., cardiac devices) would strengthen the generalizability of these findings. Second, the morphing strategy utilized a linear transformation, which does not capture non-linear anatomical deformations (e.g., skin folds) associated with extreme obesity. Additionally, trajectories were adapted to each model's anatomy to ensure realistic tissue routing, introducing minor geometric variation across models. Furthermore, tail-statistic endpoints such as $p_{95}$ and derived Maximum Allowable $B_1^+$ were estimated from $n = 50$ trajectories per model, which limits the precision of bootstrap confidence intervals, particularly for the $B_1^+$ equivalence assessment, where the nonlinear $B_1^+$ and $\Delta T$ relationship amplifies sampling variability in the upper tail. Larger trajectory sets would strengthen equivalence conclusions at the endpoint level.

\section{Conclusion}

This study establishes cross-platform equivalence between FDTD (Sim4Life) and FEM (ANSYS HFSS) implementations of the ISO/TS~10974 Tier~3 transfer function workflow for MRI safety assessment of active implantable medical devices. Despite differences in meshing strategies and numerical methods, both platforms produced statistically equivalent predictions of RF-induced heating and Maximum Allowable $B_1^+$ limits for standard male anatomical models, with $p_{95}(\Delta T)$ differences falling within the pre-specified equivalence margin of $\pm2.56~^\circ$C.

Anatomical phenotype was the dominant source of variability in heating predictions. The obese male model (Fats) required a 31\% reduction in safe $B_1^+$ exposure limits compared to standard models, a disparity that cannot be captured through dielectric property variation alone ($p_{95}$ shift $< 2~^\circ$C across $\pm50\%$ tissue property variation). This BMI-driven risk elevation was independently confirmed in female anatomy: Ella (BMI~30) showed a $p_{95}$ of 10.3$~^\circ$C and required a 19\% reduction in safe $B_1^+$ relative to the standard female reference (Ella), reinforcing BMI as the dominant variable across sexes. In contrast, sex-specific anatomy at standard BMI produced comparable heating distributions between Duke and Ella ($p = 0.22$), with overlapping Maximum Allowable $B_1^+$ limits.

Geometric morphing of an accessible standard model approximated the high-risk tail of the native obese model (Morphed HBM: 2.06~$\mu$T vs.\ Fats: 2.06~$\mu$T), offering a practical, platform-agnostic surrogate when native high-BMI models are unavailable. Homogenized variants of the morphed model consistently underestimated worst-case risk, confirming that tissue heterogeneity must be preserved for accurate surrogate predictions.

These findings suggest that safety assessments relying solely on standard-BMI anatomical models may underestimate RF heating risk in obese patients regardless of sex. Incorporating anatomical diversity, through native high-BMI models or validated morphing surrogates, would improve safety-limit estimation for high-BMI patients

\section*{Acknowledgment}
The authors acknowledge ZMT Zurich MedTech AG, Zurich, Switzerland, for providing an academic license for the Sim4Life simulation platform used in this study.

\bibliographystyle{IEEEtran}
\bibliography{references}

\end{document}